\documentclass[epj]{svjour}
\usepackage{graphics}
\usepackage{amsmath}

\begin{document}
\title{Optical Manipulation of Long-range Interactions at the 3s+3p Asymptote of Na$_2$}
\author{Chr.~Samuelis\inst{1}, St.~Falke\inst{1}, T.~Laue\inst{1}, P.~Pellegrini \inst{2}, 
O.~Dulieu\inst{2}, H.~Kn\"ockel\inst{1}, \and E.
Tiemann\inst{1}\thanks{\emph{email:}
tiemann@iqo.uni-hannover.de}%
}                     
%
\institute{Institut f\"ur Quantenoptik, Universit\"at Hannover, Welfengarten
1, 30167 Hannover, Germany  \and Laboratoire Aim\'e Cotton, CNRS, Campus
d'Orsay, B\^atiment 505, 91405 Orsay Cedex, France} 
\date{Received: date / Revised version: date}
%
\abstract{
We investigate the influence of a laser field, which is near-resonant to the atomic sodium \mbox{3$^2$P$_{1/2}$ $\rightarrow$ 3$^2$D$_{3/2}$} transition, on the last bound levels of the  A$^1\Sigma^+_u$ state in Na$_2$. In a molecular beam experiment level shifts  up to \mbox{$\approx$ 100 MHz} and light induced line broadenings were observed using an optical double resonance excitation scheme. Moreover, the coupling laser can reduce the number of bound levels of the A~state by one or more units, which effectively means that in the picture of a collision of a 3$^2$S$_{1/2}$ and a  3$^2$P$_{1/2}$-atom the scattering phase is altered by more than $\pi$. The observed effects are interpreted as light induced couplings of the A$^1\Sigma^+_u$ state, which correlates to the \mbox{3s$_{1/2}$+3p$_{1/2}$} asymptote, to the  $4^1\Sigma^+_g$ and $2^1\Pi_g$~states at the  \mbox{3s$_{1/2}$+3d$_{3/2}$} asymptote.  We performed multi-channel calculations, applying the mapped Fourier grid method, which reproduce our experimentally observed level shifts well.
} 
\authorrunning{Chr. Samuelis {\it et al.}}
\titlerunning{Optical manipulation of long-range interactions at the 3s+3p asymptote of Na$_2$}
\maketitle
\section{Introduction}
Due to the fast progress in cooling and trapping of ultracold atoms and the achievement of 
Bose-Einstein condensation \cite{An95,Da95,Bra95,Fri98,Ro01,Pe01,Weber02}, the interest in a detailed 
knowledge of ultracold collisional properties has increased \cite{Ti96}. They are often described 
by the s-wave scattering length $a_s$, which is correlated to the scattering wave function at 
vanishing collisional energy. It has been shown \cite{Cru99,Elbs99,Samu2001} that 
this important physical quantity can be derived from spectroscopy of the vibrational levels and 
resonances around the atomic asymptote. But the scattering length is difficult to derive purely on theoretical grounds because it does not only depend on the long-range 
part of the molecular ground state interaction potentials but also on the accumulated phase from the inner part of the potential.

Beside the knowledge of cold collision properties, the possibilities of their manipulation are 
central points of interest. It has been shown in Na-Na, Rb-Rb, and recently also in Cs-Cs ultracold 
collisions that the sign of the scattering length can be altered in the vicinity of a magnetic 
field induced Feshbach resonance by varying the magnetic field strength \cite{In98,Cor00}. This 
offers a wide range of experiments on the dynamics in ultracold ensembles and BEC and a further 
understanding of two and three particle interaction. Unfortunately, magnetic tuning of two particle 
interaction is not applicable in typical magnetic traps because the trapping field 
conditions are affected by the field strength needed according to the desired tuning of the 
collision conditions.

An alternative approach for the manipulation of cold collisions has been developed in \cite{Ma98}, 
where DC electric fields are proposed to control diatomic collisions at ultralow temperatures.

A further promising proposal is the use of near resonant light fields to influence the scattering 
length $a_s$ \cite{Fed96,bohn97,kok01}. This technique offers the possibility to manipulate the two-particle 
interaction by coherent light fields. First experiments with photoassociation \cite{Fat00} show 
that it is possible to observe the optical analogue to the magnetic field induced Feshbach resonances
if a laser with a frequency close to a photoassociation transition is focused into a magneto optical 
trap. The study of an optically induced Feshbach resonance is important for possible control of 
condensate dynamics.
  
In this paper we follow another approach to determine the effect of near resonant laser light on 
cold atomic collisions. We use the high resolution of Doppler free spectroscopy on a molecular 
beam to investigate the effect of a laser field that is near resonant to the atomic sodium 
\mbox{3$^2$P$_{1/2}\rightarrow$3$^2$D$_{3/2}$} transition. Thus, this laser  induces a coupling 
of the A$^1\Sigma^+_u$ state at the \mbox{3s$_{1/2}$+3p$_{1/2}$} asymptote to states at the 
\mbox{3s$_{1/2}$+3d$_{3/2}$} asymptote ($4^1\Sigma^+_g$, $2^1\Pi_g$) with their vibrational 
and continuum manifold. This results in energy shifts of asymptotic rovibrational energy levels 
with internuclear separations of about 100~\AA\ in all coupled states modeling atomic pairs with low relative kinetic energy. We extract quantitative information on light induced coupling from the 
induced level shifts. A similar laser controlled manipulation of the asymptotic ground state 
potential can be used to investigate schemes to modify the ground state s-wave scattering length 
$a_s$. The advantage of the experiment at the  \mbox{3s$_{1/2}$+3p$_{1/2}$} asymptote presented 
below is the simpler detection via direct laser-induced fluorescence, which is not possible 
in the case of the ground state asymptote.

This article is organized as follows: First, an outline of our experimental setup and the excitation 
scheme is given (Section~\ref{exp}). In the following Section~\ref{fitdata} we focus 
on the extraction of quantitative data from the observed line profiles. Experimental results are 
presented in Section~\ref{expdata}. 
A theoretical model that simulates the investigated light 
induced energy shifts and line broadenings utilizing a coupled-channel
calculation including six channels is developed in Section~\ref{theo} in the 
framework of the Mapped Fourier
Grid Hamiltonian (MFGH) method involving a complex potential
\cite{Koko99,pell02}.
In the concluding section we compare the results of the simulation 
with the experiment and give also prospects for further experiments and extended simulations.

\section{Experiment}
\label{exp}
\begin{figure}
\resizebox{0.45\textwidth}{!}{\includegraphics{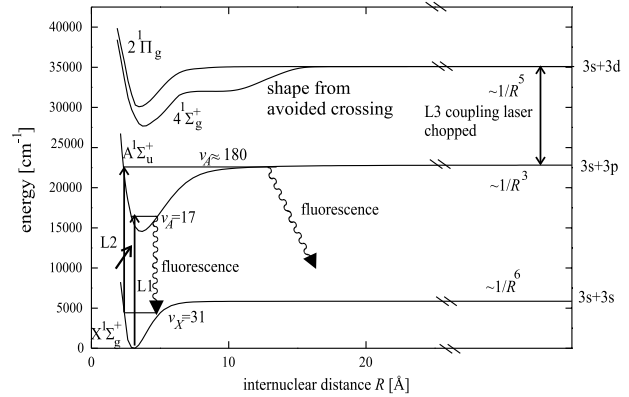}}
\caption{Potential scheme of Na$_2$. Laser L1 transfers population to a desired vibrational level 
\mbox{($v_X=31$),} so that the \mbox{3s+3p} asymptote can be reached via the second 
laser L2. The oblique arrow indicates that this laser is scanned during the experiment. A coupling of asymptotic A state levels to the \mbox{3s+3p} asymptote is induced via 
the coupling laser L3. The fluorescence from the A state is detected with a lock-in detector 
synchronized to the chopper frequency of laser L3.}
\label{fig1}
\end{figure}

\begin{figure}
\resizebox{0.45\textwidth}{!}{\includegraphics{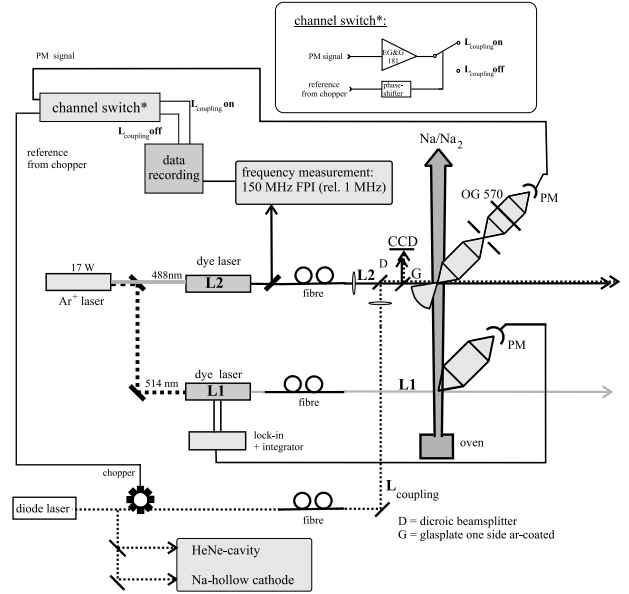}}
\caption{Experimental Setup. A well-collimated molecular beam is crossed by three laser beams. Laser 
L1 is stabilized on the molecular fluorescence in the first interaction zone, while the fluorescence 
in the second interaction zone is recorded. The perturbing laser L3 is locked on a stabilized cavity. 
The signal from the second interaction zone is processed with a channel switch to record a manipulated 
(L3~on) and unmanipulated (L3~off) spectrum, simultaneously.}
\label{fig2}
\end{figure}

We briefly review the experimental setup and focus on the changes made in comparison to \cite{Tie96} 
were we studied asymptotic sodium A~state levels within our molecular beam experiment. For the 
investigation of light induced energy shifts below the asymptote of the A~state, we start from 
molecules in a well collimated beam (molecular velocity of \mbox{$\approx 1000\:$m/s)} which are 
initially in the lowest vibrational levels \mbox{($v_X=\:$0, 1)} of the $X^1\Sigma^+_g$~ground state 
(see \mbox{Figure \ref{fig1}}). In a first interaction zone (\mbox{Figure \ref{fig2}}) a Franck-Condon 
pumping step is applied with laser L1 to create population in thermally unpopulated vibrational 
levels  \mbox{$v_X\approx\:$31} of the ground state. For this purpose a dye laser 
(sulforhodamine b) operating at about 615 nm is used. Its frequency is stabilized on the maximum of 
fluorescence of the selected molecular  transition.

Starting from such vibrationally excited ground state levels the A~state asymptote is reached with a 
tunable dye laser L2 at 532~nm operating with coumarin~6. It is applied in a second interaction zone 
about 0.35~m downstream from the first zone. The fluorescence from the excited levels is monitored 
by a photomultiplier. An OG~570 color glass filter suppresses the scattered laser light from L2.

For the manipulation of bound levels close to the A~state asymptote a third laser L3 is used. It 
is an extended cavity diode laser in Littrow configuration operating near the atomic sodium 
\mbox{${\rm 3^2P_{1/2}\rightarrow 3^2D_{3/2}}$} transition at \mbox{818 nm.}  It is overlapped with L2 
by a dichroitic beam splitter and focused into the second interaction zone. In that way a coupling of 
the \mbox {A${\rm ^1\Sigma^+_u}$} state to the 4${\rm ^1\Sigma^+_g}$ and the $2{\rm ^1\Pi_g}$~state 
at the  \mbox{3s+3d} asymptote is created. The waist radius is \mbox{$\approx 80\mu$m} for L2 and 
about \mbox{$\approx 100\mu$m} for the laser L3. Overlap and diameter of both beams are controlled 
with a CCD camera. Both laser beams are linearly polarized, and their polarization axes are parallel 
to the molecular beam.

For the determination of the atomic transition frequency \mbox{${\rm 3^2P_{1/2}\rightarrow 3^2D_{3/2}}$} 
we apply optogalvanic spectros\-copy in a conventional sodium hollow cathode lamp as reference. By this means a Doppler 
broadened spectrum of the atomic transition can be obtained. The position of the peak can be determined 
with an uncertainty of about 30~MHz. For a good long-term frequency stability of L3 
it is stabilized to a 150~MHz marker cavity that itself is locked to an iodine stabilized HeNe laser 
\cite{PMT00}. Long term drifts of L3 less than 1~MHz/h are achieved in this way. The cavity is 
locked to the HeNe laser using an acousto-optic modulator (AOM) which allows interpolation between 
the 150~MHz markers. Hence, we are able to determine the relative detuning with an absolute uncertainty 
in the order of 1~MHz as long as the cavity is locked to the HeNe laser (see Section~\ref{expdata}).

The effect of L3 on the energetic position of asymptotic levels in the A~state is observed by a switching 
technique: L3 is modulated with a mechanical chopper at a frequency of about 1~kHz. The current of the 
photomultiplier in the second interaction zone is amplified with a fast current amplifier and then fed 
into an electronic switch, triggered by the chopper signal. It divides the signal into two output 
channels: The first channel is correlated to \lq L3 off\rq, the second to \lq L3 on\rq\, if the switching 
phase is adjusted properly.

\begin{figure}
\resizebox{0.45\textwidth}{!}{\includegraphics{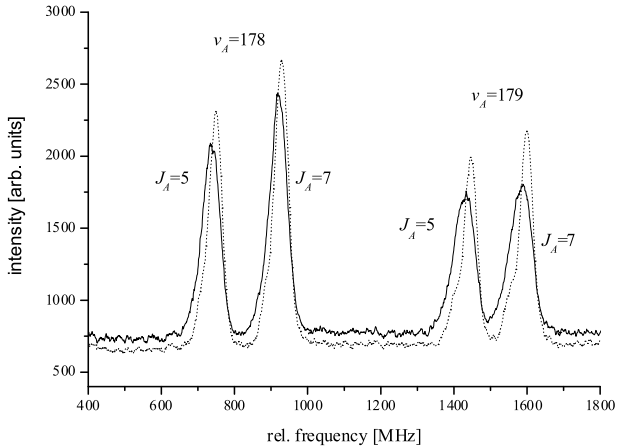}}
\caption{A spectrum $\approx 2.5\:$GHz below the asymptote of the A~state. L1 was stabilized on the 
\mbox{R(6)(17-0)} transition, and L2 is tuned across the P(6)/R(6) transitions $v_X=31\rightarrow 
v_A=178,179$. The dotted line is a signal with the coupling laser switched off, while the solid 
one corresponds to the situation when the coupling laser is on ($I_{\rm L3}\approx 74\:$W/cm$^2$). 
The background results from scattered light of L2. The additional offset in the solid spectrum is 
due to additional scattered light from the coupling laser. This laser is stabilized 
in frequency with a detuning of 75 MHz to blue with respect to the atomic sodium 
\mbox{3$^2$P$_{1/2}\rightarrow$ 3$^2$D$_{3/2}$} transition. For both vibrational levels the line 
profiles of the P and R line are asymmetric. Each line has a shoulder on the low frequency (left) 
side originating from underlying hyperfine structure.}
\label{fig3}
\end{figure}

A typical recorded spectrum, tuning laser L2 while keeping L1 and L3  fixed, is shown in 
\mbox{Figure \ref{fig3}}. Several scans are averaged to improve the signal-to-noise ratio of the 
recording. For the presented scan the total integration time per data point was 1s. The Franck-Condon 
pumping laser L1 is usually stabilized to the A-X \mbox{R(6)~(17-0)} transition. The population in $v_A\:$=17, 
$J_A\:$=7 decays to $J_X\:$=6 and 8 preferentially of vibrational levels around $v_X\:$=31. 
Starting from $v_{X}\:$=31, $J_{X}\:$=6, L2 stimulates the transitions P(6) and R(6) to 
the asymptotic vibrational levels $v_A\:$=178 and 179. The spectrum for every $v_A$ thus 
consists always of two angular momentum components $J_A=J_X\pm1$ for each $v_A$, to which 
we refer as P line ($J_A=J_X-1$) and R line ($J_A=J_X+1$) in the following. The spectrum is 
shifted to red (lower frequency) by laser L3, the lines are broader and show asymmetry.

\section{Extraction of Line Shifts and Line Broadenings}
\label{fitdata}
In this section we explain how we extract information on the line shift~$\Delta$ between the manipulated 
and the unmanipulated trace in a spectrum like in \mbox{Figure \ref{fig3}}. We determine the induced 
light shift by applying a line shape fit. We deal with the two channels --- with and without the coupling 
laser L3 --- simultaneously because the two traces can be described by several common line shape parameters 
and few additional parameters describing the line shift. It is useful to concentrate on simultaneous simulation 
of P/R doublets because the two lines overlap for large induced line shifts. The line broadening 
is determined within the same line shape fit.

For the description of the spectrum several hyperfine components underneath each rotational line have 
to be taken into account: Due to nuclear spin statistics both P and R line for an even $J_X$ have 
contributions from nuclear spins $I=0,2$. For the latter different orientations with respect to 
$J_A$ appear, which results to 1+5 hyperfine components. For the electric dipole transitions 
$I$ and $M_I$ remain unchanged to good approximation. Each contributing hyperfine component is described 
by a standard profile 
\begin{equation}
\label{eqn1} 
{{{\rm A}\left[{\rm A_0}, \nu,\nu_0\right]}=\frac{\rm A_0}{1+\alpha X^2+(1-\beta-\alpha)X^4+\beta^6 X^6}}
\end{equation}
with
\begin{equation}
{{X}={\frac{\nu-\nu_0}{\mathrm{HWHM}}}}.
\end{equation}

The parameters $\alpha$, $\beta$ and HWHM (half width at half maximum) are equal for all lines contributing 
to the unshifted spectrum. $\alpha$ and $\beta$ are line profile parameters, which depend on the contribution 
of the Lorentzian ($\alpha=1, \beta=0$) and residual Gaussian shapes to the line shape. They are determined 
by the collimation ratio of the experimental apparatus (1:1000) and possibly by frequency jitter. For the profile fits presented here, 
$\alpha$ has been kept fixed at 1 while $\beta$ is variable. 

The fit results show that the observed line profiles can be simulated using the scalar and the tensorial 
nuclear spin-spin interactions. The center frequency $\nu_0$ and the amplitude of each line component 
thus does not only depend on the rotational quantum number $J_A$ and  the nuclear spin $I$, but also 
on the total molecular spin $F=J+I$. The total contribution diagonal in $J$ of the hyperfine interaction 
to the line position is \cite{Bro78}:
\begin{equation}
\begin{split}
\label{Hyperfein}
E_{\rm HFS}=&\langle \Omega=0 J I^{'} F| H_{\rm HFS}|\Omega=0 J I F\rangle \\
=&\frac{\delta}{2}\delta_{I I^{'}}
\bigg[ I(I+1)-2 i(i+1)  \bigg]\\
&+ d \hspace{5pt} i(i+1)(2i+1)(-1)^{I^{'}+F+1}\\&\times\bigg[30(2I+1)(2I^{'}+1)\bigg]^{1/2}\\
&\times(2J+1) 
\left(     
\begin{array}{ccc}
J&J&2\\
0&0&0
\end{array}
\right)
\left\{     
\begin{array}{ccc}
F&I^{'}&J\\
2&J&I
\end{array}
\right\}
\left\{     
\begin{array}{ccc}
i&i&1\\
i&i&1\\
I&I^{'}&2
\end{array}
\right\}.
\end{split}
\end{equation}
$\vec{I}$=$\vec{I^{'}}$=$\vec{i}_1$+$\vec{i}_2$ is the total nuclear spin, 
$i=i_1=i_2=3/2$ is the nuclear spin of a single sodium atom and
Wigner's symbols are used. The first term in equation (\ref{Hyperfein}) is the scalar part 
of the hyperfine interaction while the second one is the tensorial contribution of the hyperfine energy.  
The $\delta$-parameter for the scalar and the $d$-parameter for the tensorial part depend on the states 
mixed by the hyperfine interaction. At the 3s+3p asymptote a large variety of states is coupled to the 
A~state, so that the functional dependence of the ratio of $\delta$ and $d$ does not have a simple analytic 
form. Thus, for the P/R line we let both parameters vary in the fit. The quadrupole and magnetic rotation 
hyperfine interactions are negligible in the present case of asymptotic levels.

For the profile simulations, we assume that all hyperfine components for a single $J$ resulting from equation 
(\ref{Hyperfein}) are of equal intensity and thus we need two parameters for the description of the 
intensities of each P/R doublet. This assumption would be not sufficiently good for small angular momenta $J,F$. 
For the positions of P and R lines $\nu_0=\nu_v+B_vJ_A(J_A+1)+E_{\rm HFS}$ is used. For every $v_A$ 
the unperturbed profile is composed by (for even $J_X$)
\begin{eqnarray}
\label{eqUn}
U\left(\nu\right)&=& U_{\rm off} + \sum_{J_A=J_X\pm 1}\hspace{1ex}\sum_{
I=0,2}\\
&&\sum_{F=\left|J_A-I\right|}^{J_A+I} {\rm A}\big[{\rm A_0}\left(I,F,J_A\right),
\nu,\nu_0\left(I,F,J_A\right) \big], \nonumber
\end{eqnarray}
with $U_{\rm off}$ to account for a background offset in the experimental trace.

For the simulation of the recordings with the coupling light field on, we use the same parameters $\alpha$ 
and $\beta$ but a different HWHM. It can be larger than without coupling laser due to predissociation broadening 
induced by the optical coupling to the upper states ($4^1\Sigma^+_g$, $2^1\Pi_g$, see discussion below). 

The offset of the trace recorded with coupling laser $S_{\rm off}$ differs from the offset $U_{\rm off}$ of the uncoupled trace as the 
detector sees some additional scattered light originating from the coupling laser. We introduce an additional
parameter for the description of the absolute intensities of both P and R line of the shifted spectrum.

A coupling laser field causes the splitting of a line at $\nu_0$ into two Autler-Townes components 
at $\nu_0\pm\Delta$.  The line shift $\Delta$ and the relative weight of the Autler-Townes 
components of each line depend on the frequency of the coupling laser and on the Rabi frequencies of the 
coupling transitions. For  P and R line the ratio of the intensities of the two Autler-Townes components are fitting parameters 
and are later called $q_{J_A}$. Due to reasons discussed below the shifted doublet is composed of various 
lines with different shifts $\Delta$.

The geometries of the laser beams L2 an L3 need to be considered for a proper simulation: The detection 
probability is proportional to the local intensity of L2, $I_{\rm L2}(r)$, and as the dipole coupling is $M$ 
dependent, the induced shift is a function of $M$ and of the local intensity of L3, $I_{\rm L3}(r)$.

As the polarization vector of L2 and of L3 are defined in the laboratory frame and the transition moments of 
the molecules are given in the molecular frame a transformation between those two systems has to be performed.  
We assume that the projection $M$ of the angular momentum $J$ onto to the laser polarization vector 
remains unchanged in the second interaction zone, and that all $M$ are equally populated. The latter 
assumption is justified as we start from a thermal molecular beam and  the light of L1 is unpolarized.

As P, R, and Q (state $2^1\Pi_g$) coupling contribute to the induced shift of a level for a specific $M$, the $M$-dependence 
is not the regular parabolic dependency as for a single P, Q, or R coupling like $M^2$  or $J^2-M^2$ that 
results from the direction cosine matrices. We derived from a simulation with diabatic, rotationally 
corrected potentials and {\it ab initio} coupling strengths (see Section~\ref{theo}) that for 
large internuclear distances the  shift $\Delta$ resulting from P, Q, and R coupling can be factorized 
into a function parabolic in $M$ and linear in $I_{\rm L3}$:
\begin{equation}
{{\Delta\left(M,I_3\right)}\sim{I_{\rm L3} \left[1- b \left(\frac{M}{J_A}\right)^2\right]}}
\label{Mdependence}
\end{equation}
We derived $b=0.7$ in our simulation. This result of potential shift calculations 
at large internuclear distance is applied to asymptotic levels because in such cases the main contribution of the overlap integral of 
the vibrational wave functions results from large internuclear separations.

For the geometric model of the laser intensity distribution we assume Gaussian \mbox{TEM$_{00}$} laser 
modes. Although the beams L2 and L3 are focused (waist  $w_0^{2,3}$) into the second interaction zone, 
we can assume that we have plane waves because the Rayleigh ranges $z_R^{2,3}$ are larger than the radius 
of the detection zone $\rho$ ($w_0^i\ll \rho \ll z_R^{i}$). With good approximation we can assume that 
the molecular density is constant in the observed zone. Taking the cylindrical symmetry of the 
laser fields into account, the description of the space dependent interaction can be reduced to one 
coordinate: the radial distance~$r$ from the common axis of the laser beams L2 and L3. We define as an 
approximation to a Gaussian profile a distribution of a discrete number $p_{\max}$ of equi-thick hollow 
cylinders. In each of them we assume a constant intensity $I_{\rm L2}^p$ and $I_{\rm L3}^p$ representing 
the average intensity of L2 and L3 in the cylinder~$p$ where $p$ counts from the center to the outer range.

With this model the profile of one Autler-Townes component of one rotational line $J_X\rightarrow J_A$ with 
an unperturbed transition frequency $\nu_0$ can be described as follows:
\begin{eqnarray}
\label{eqT}
&&{\rm T^\pm}\left[\nu,\nu_0,J_A,{\rm A_0}\right] \sim \sum_{M=-J_X}^{J_X}
\left( 
\begin{array}{ccc}
J_A & 1 & J_X \\
-M & 0 & M
\end{array}
\right)^2\\
&&\times \sum_{p=1}^{p_{\max}} \frac{2p-1}{p_{\max}^2}\;
I_{\rm L2}^p\;{\rm A} \big[ \nu, \nu_0 \pm \Delta (M,I_{\rm L3}^p),{\rm A_0}\big] \nonumber.
\end{eqnarray}
The Wigner 3j-symbol gives the relative detection probability for a single $M$ in the unsaturated case for 
L2 and $\frac{2p-1}{p_{\max}^2}$ is the statistical weight for the hollow cylinder~$p$ of molecules. Like 
for the unshifted spectra described by equation (\ref{eqUn}), the summation over all  contributing lines 
needs to be performed for the shifted spectra, too:
\begin{eqnarray}
&&S\left(\nu\right)=S_{\rm off} + \sum\limits_{J_A=J_X\pm 1}\hspace{1ex}\sum\limits_{I=0,2}
\hspace{1ex}\sum\limits_{F=\left|J_A-I\right|}^{J_A+I} \\
&& \times \bigg[  q_{J_A}^{\pm} {\rm T^+}\big[\nu,\nu_0\left(I,F,J_A\right),J_A,{\rm A_0}
\left(I,F,J_A\right)\big] +   \nonumber\\
&& \hspace{3ex} \left(1-q_{J_A}^{\pm}\right)  {\rm T^-}\big[\nu,\nu_0\left(I,F,J_A\right),J_A,{\rm A_0}
\left(I,F,J_A\right)\big]
\bigg] \nonumber . 
\end{eqnarray}
The intensity ratios of the Autler-Townes components $q_{J_A}^{\pm}$ are limited by $0 \le q_{J_A}^{\pm} \le 1$ and 
are either almost 0 or almost 1 because the detuning of L3 from the coupling transitions is rather large 
compared to the Rabi frequency (see above).

In the following the derived parameter of the level shift corresponds to the maximal shift, i.e., the shift for 
$M=0$ molecules at the beam center, $\Delta(0,I_{\rm L3})$ in equation (\ref{Mdependence}).

\begin{figure}
\resizebox{0.45\textwidth}{!}{\includegraphics{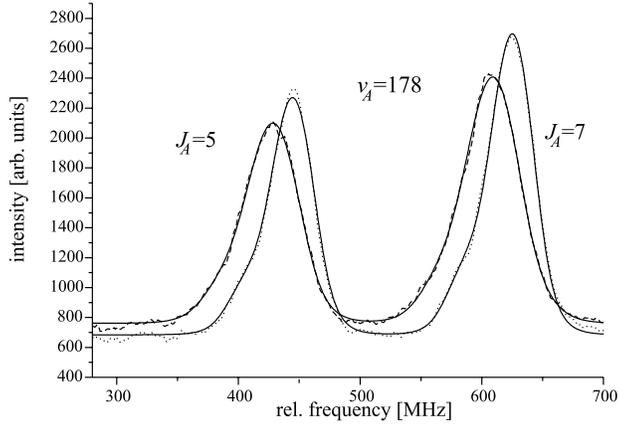}}
\caption{Experimental and simulated spectrum of $v_A=178$, $J_A=5,7$. The dotted trace is the unshifted 
experimental spectrum. The dashed trace is the shifted experimental one (coupling laser on the atomic 
\mbox{3$^2$P$_{1/2}\rightarrow~3^2$D$_{3/2}$} resonance, $I_{\rm L3}=\:$74 W/cm$^2$). The two solid traces 
correspond to the simulation of the spectra as explained in the text.}
\label{fig4}
\end{figure}

\mbox{Figure \ref{fig4}} shows the result of a fit to an experimental spectrum for $v_A=178$. The 
experimental traces are plotted in  dotted (unshifted) and dashed style (shifted). L3 was resonant 
to the \mbox{3$^2$P$_{1/2}\rightarrow$3$^2$D$_{3/2}$} transition and caused an intensity of $I_{\rm L3}= 74 \:$W/cm$^2$. The 
criterion for the quality of the fit is the sum of squared difference between experimental and fitted points. 
The agreement between experimental and fitted line profiles is within the noise interval. The line shift 
obtained from the fit is -28.9 MHz for $J_A\:$=5 and -29.7~MHz for $J_A\:=$7 in the above case. The uncertainty 
of the shifts can be estimated by varying the starting conditions of the fit and is determined to be 
less than 3~MHz here.

\section{Experimental Results}
\label{expdata}
\begin{figure}
\resizebox{0.45\textwidth}{!}{\includegraphics{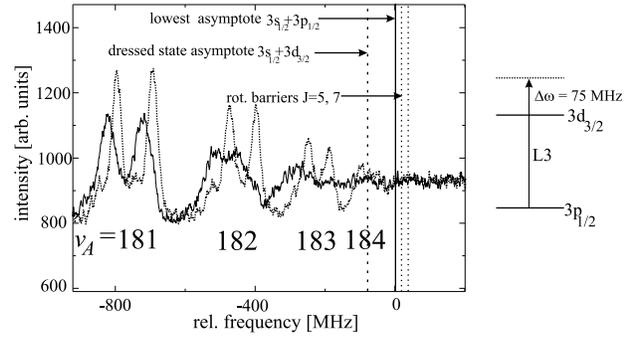}}
\caption{Effect of the optical manipulation on the last vibrational levels of the A state. The coupling 
laser is 75(30)~MHz blue detuned with respect the atomic \mbox{3$^2$P$_{1/2}\rightarrow$3$^2$D$_{3/2}$} 
transition. The intensity is 43 W/cm$^2$. The dotted spectrum is recorded in absence of the coupling laser 
while the solid one is recorded with coupling laser switched on.}
\label{fig5}
\end{figure}

The dependence of the light induced energy shifts of $v_A\:=173,\dots,182,$ $J_A\:$=5,7 on 
power and detuning of the coupling laser L3 has been investigated. First we will focus on the effect on 
the last bound levels directly below the A state asymptote. Figure~\ref{fig5} shows a spectrum of the 
levels $v_A=181$ to 184. The trace plotted in dotted style is recorded without the coupling laser while 
the second trace, plotted in solid style, is recorded in presence of the coupling laser field. The intensity 
of the coupling laser was set to $I_{\rm L3}=43$ ${\rm W/cm^2}$ and the  frequency was stabilized 75(30)~MHz 
blue detuned with respect to the atomic \mbox{3$^2$P$_{1/2}\rightarrow$3$^2$D$_{3/2}$} transition. All levels 
are shifted to lower frequencies or energies, note the position of the potential asymptote. Additionally, 
the line profiles of the levels $v_A\:=182$ to 184 have remarkably changed: The lines are  broader than 
without the coupling laser. The P and the R line start to overlap. The dissociation continuum has been shifted, \mbox{$v_{A}=184$} disappears.

\begin{figure}
\resizebox{0.45\textwidth}{!}{\includegraphics{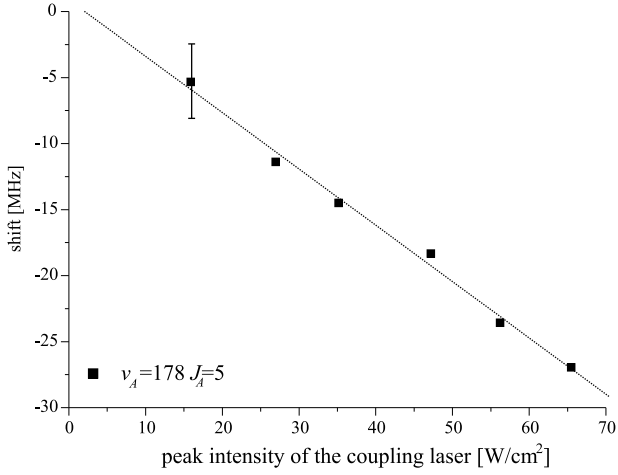}}
\caption{Dependence of the light shift of $v_A\:$=178,~$J_A\:$=5 on the intensity 
of the coupling laser. Without coupling laser this level is bound by 2646~MHz. The frequency of the coupling laser is stabilized 75(30)~MHz blue detuned from the atomic \mbox{3$^2$P$_{1/2}\rightarrow$3$^2$D$_{3/2}$} 
transition. The absolute uncertainty of the coupling laser intensity is 50\% and originates from the resolution 
of the beam view camera utilized for the determination the laser beam diameters. The relative uncertainty between different 
measurements is less than 5\%.}
\label{fig6}
\end{figure}

In Figure~\ref{fig6} the variation of the observed shifts of $v_A\:$=178, $J_A\:$=5 as a 
function of the intensity of the coupling laser is shown. The laser frequency of L3 was kept fixed 75(30)~MHz 
blue detuned as before, but the intensity was varied from 15 to 67~W/cm$^2$. This corresponds to energy shifts 
in the central region of the beam up to 26.9(30)~MHz. The dependence of the shift on the laser intensity is 
linear within the experimental accuracy.

The error limits for the absolute intensity result mainly from the uncertainty in the determination of the focus 
diameter of the coupling laser beam. We use for this purpose a CCD camera with a pixel size of 23$\mu$m 
$\times$ 27$\mu$m, on which we project a 10\% reflex of each laser beam of the second interaction zone. 
The CCD and the second interaction zone are equi-distant from the beam splitter, so that the beam diameter 
measured with the CCD is the same as the focus diameter in the second interaction zone and can be 
derived with an absolute uncertainty of one pixel size of the CCD chip which results in the 50\% uncertainty 
for the determination of the coupling laser intensity. The relative change of the intensity in 
\mbox{Figure \ref{fig5}} is known with an uncertainty of 5\%, limited by the uncertainty of the power meter.

As an example for the dependence of level shifts on the detuning of the coupling laser experimental shifts 
of $v_A\:$=179,~$J_A\:$=7 are plotted in Figure~\ref{fig10}. The reference frequency 
for the detuning is the energy difference $\delta \omega$ from $v_A\:$=179,~$J_A\:$=7 to the 
\mbox{3s$_{1/2}$+3d$_{3/2}$} asymptote. This is reasonable because in the region of the outer turning point 
of the \mbox{A state} vibrational wave function (where the coupling is induced) the potential of the upper 
\mbox{3s+3d} asymptote has almost reached its asymptotic value. This is due to the $1/R^5$-dependence from a 
quadrupole-quadrupole coupling of a s-electron and a d-electron and is in contrast to the $1/R^3$-behavior 
of the resonant dipole-dipole interaction of the A state potential correlated to the s+p asymptote. For 
the series of measurements in Figure~\ref{fig10} the laser intensity of the coupling laser was held fixed at 74 W/cm$^2$.
The circles in Figure~\ref{fig10} are results from simulations, which we will describe in the following section.

\section{Theoretical Interpretation}
\label{theo}

The observed level shifts originate from the laser-induced dipole coupling of 
the A$^1\Sigma^+_u$~state to the 4$^1\Sigma^+_g$ and the 2$^1\Pi_g$~states, both
correlated to the 3s+3d asymptote (Figure~\ref{fig1}). We use a dressed
molecule picture similar to \cite{Vat01} for the simulation of the coupled 
system. After eliminating the explicit time dependence from the theoretical 
model, a mapped Fourier grid method \cite{Koko99} is used to determine line 
positions and light induced line broadenings.

\begin{figure}
\resizebox{0.45\textwidth}{!}{\includegraphics{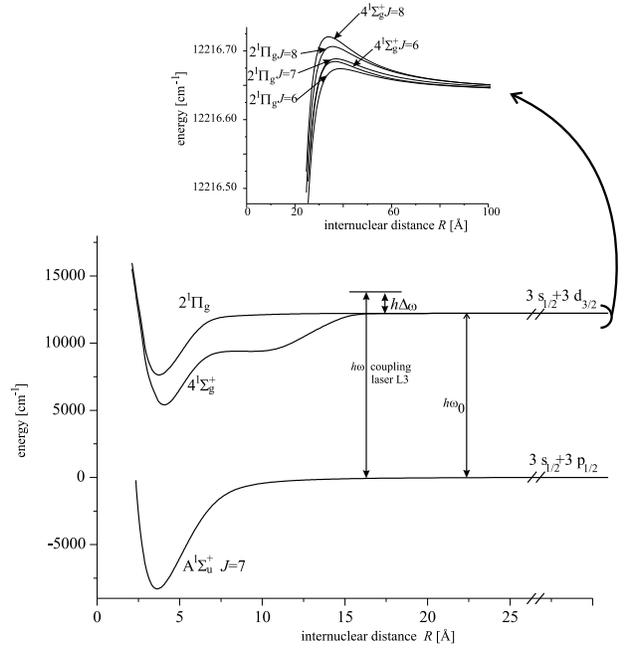}}
\caption{Potential scheme of the dipole coupling of A$^1\Sigma^+_u$~state to the
 4$^1\Sigma^+_g$ and the 2$^1\Pi_g$~states. Dipole selection rules allow P and R
coupling to the 4$^1\Sigma^+_g$-state, and P, Q, and R coupling to the 
2$^1\Pi_g$-state. The inset shows the rotational barriers at the $3s+3d$
asymptote
for $J_A=7$.}
\label{fig7}
\end{figure}

For the correct description of the dipole coupling we have to take care of 
the e/f symmetries of the involved rotational levels and the selection rules for
an optically induced dipole coupling. Thus, the A state can only couple to the 
$4 ^1\Sigma^+_g$~state on a P and R line, while there is  P, Q, and R coupling 
to the $2 ^1\Pi_g$-state. For a proper symmetrization, the latter state usually
is expressed as a linear combination \mbox{ $|2 ^1\Pi_g \rangle = 
\frac{1}{\sqrt{2}} \left( |\Omega=1\rangle \pm (-1)^J|\Omega=-1\rangle 
\right)$}. The plus sign gives the e component of the $\Pi$-state that can only
be reached by P and R couplings, while the minus is the f component which
couples 
to the A~state by a Q line. For each $M$ (projection of $J$ onto space fixed 
axis neglecting hyperfine structure), a coupled system can be calculated. Thus, 
the system to be solved  consists of six (except five for $M$=0, where the Q coupling 
for \mbox{$\Sigma\rightarrow\Pi$} vanishes) coupled channels, as 
illustrated in Figure~\ref{fig7}.

The coupled system is described by a six component wave function. For a simple
notation we use $G$, $\Sigma$, and $\Pi$ instead of A$^1\Sigma^+_u$, 
4$^1\Sigma^+_g$, and 2$^1\Pi_g$ for the electronic part of the wave functions.
With Born-Oppenheimer basis functions, the coupled channel wave function thus
can be expressed as
\begin{equation}
\begin{split}
\vec \Psi (R,t)=&
\left( 
\begin{array}{r}
\Psi_{G}^{(J)}(R, \vec{r}_k, t)\\
\Psi_{\Sigma}^{(J-1)}(R, \vec{r}_k,t)\\
\Psi_{\Sigma}^{(J+1)}(R, \vec{r}_k,t)\\
\Psi_{\Pi}^{(J-1)}(R, \vec{r}_k,t)\\
\Psi_{\Pi}^{(J)}(R, \vec{r}_k,t)\\
\Psi_{\Pi}^{(J+1)}(R, \vec{r}_k,t)\\
\end{array}
\right)\\
=&
\left( 
\begin{array}{rcl}
\frac{1}{R} \chi_{G}(R,t) & \phi_{G}(\vec{r}_k;R) & |\xi_J\rangle\\
\frac{1}{R} \chi_{\Sigma}(R,t) & \phi_{\Sigma}(\vec{r}_k;R) & |\xi_{J-1}\rangle\\
\frac{1}{R} \chi_{\Sigma}(R,t) & \phi_{\Sigma}(\vec{r}_k;R) & |\xi_{J+1}\rangle\\
\frac{1}{R} \chi_{\Pi}(R,t) & \phi_{\Pi}(\vec{r}_k;R) & |\xi_{J-1}\rangle\\
\frac{1}{R} \chi_{\Pi}(R,t) & \phi_{\Pi}(\vec{r}_k;R) & |\xi_J\rangle\\
\frac{1}{R} \chi_{\Pi}(R,t) & \phi_{\Pi}(\vec{r}_k;R) & |\xi_{J+1}\rangle\\
\end{array}
\right),
\label{basis}
\end{split}
\end{equation}
where $R$ is the distance of the two nuclei and $\vec r_k$ are the coordinates of 
the electrons in the molecular frame. $|\xi_J\rangle$ are the basis functions
for the rotation for an angular momentum~$J$. $\phi_{G}(\vec{r}_i;R)$, $ 
\phi_{\Sigma}(\vec{r}_i;R)$, and $ \phi_{\Pi}(\vec{r}_i;R) $ are the electronic 
wave functions corresponding to the potential curves $U_{G}$, $U_{\Sigma}$, and 
$U_{\Pi}$. The functions $\chi_{G}$, $\chi_{\Sigma}$, and $\chi_{\Pi}$ describe 
the radial motion in these potentials. 

The Hamiltonian of the coupled system divides into two parts:
\begin{eqnarray}
\underline{\hat {\bf H}} = \underline{ \hat {\bf H}}^{\rm mol} + \underline{ \hat {\bf W}}.
\end{eqnarray}

The molecular part of the Hamiltonian  $\underline{\hat {\bf H}}^{\rm 
mol}$describes the field free case, whereas the second part $\underline{ \hat 
{\bf W}}$ describes the dipole coupling between the different electronic states. 
Both matrices are set up as a matrix of $k\times k$ blocks where $k$ is the 
number of coupled channels. The diagonal blocks of $\underline{\hat {\bf 
H}}^{\rm mol}$ are the matrixes $\hat {\bf H}^{\rm mol}_{i} 
\left(1\le~i\le~k\right)$, all non-diagonal blocks vanish. $\underline{ \hat 
{\bf W}}$ is set up of $\hat {\bf W}_{ij} \left(1\le~i,j\le~k\right)$. $\hat 
{\bf W}_{ij}$ represents the dipole coupling between the states of channel $i$ 
and $j$. $\hat {\bf W}_{ij}$ vanishes if $i$ and $j$ are equal or if none of 
the channels $i,j$ is $G$. Operators denoted with underscore, i.e., 
$\underline{\rm \hat {\bf H}}$, represent operators that act on all channels (in 
our case five or six  for all internuclear distances). In contrast operators
without underscore represent operators that either act on one channel only or 
describe the coupling of exactly two channels ($\hat {\bf H}^{\rm mol}_{i}$ 
resp. $\hat {\bf W}_{ij}$).

\begin{equation}
\begin{split}
&\hat {\bf H}^{\rm mol}_{i}= \hat {\bf T}_i + \hat {\bf U}_{i}\\
&+\frac{\hbar^2 
\left[J \left(J +1\right)+S \left(S +1\right)+L \left(L +1\right)-\Omega^2-\Sigma^2-\Lambda^2\right]_i}{2 \mu R^2}
\label{eqU}
\end{split}
\end{equation}
is the molecular part of the Hamiltonian. $\hat {\bf T}_i$ is the kinetic energy
operator and $ \hat {\bf U}_i$ the electronic potential energy operator of the
electronic state $i$. The third term is the rotational energy of the two nuclei 
relative to each other represented in Hund's coupling case (a). $\mu$ is the reduced 
mass of the molecule. 
 
The coupling laser L3 is assumed to be a plane and monochromatic wave 
with frequency  $\omega_L$. Thus, in dipole approximation

\begin{equation}
\hat {\bf W} \left(t\right)_{ij}= - \vec D_{i,j}(R) \cdot \vec \epsilon_L E_0 \frac{1}{2}\left(e^{i\omega_L t}+e^{-i\omega_L t}\right). 
\label{eqRWA}
\end{equation}
$\vec D_{i,j}(R)$ is the dipole moment for the transition between the electronic
states $\phi_{i}$ and $\phi_{j}$ resulting from the integration over $\vec{r}_k$ according to equation (\ref{basis}). $\vec D_{i,j}(R)$ vanishes for transitions 
between $\Sigma$- and $\Pi$-state because $g \leftrightarrow g$ couplings are 
forbidden by electric dipole selection rules. $\vec \epsilon_L$ describes the 
unity vector in the direction of the laser polarization and $E_0$ is the 
electric field strength.

With this Hamiltonian, the time-dependent Schr\"odinger equation for the six-components wave function $\vec \Psi(t)$ is:
\begin{equation}
i \hbar \frac{\partial}{\partial t }  \vec \Psi(R,t)= \left[\underline{\hat {\bf H}}^{\rm mol}+\underline{\hat {\bf W}}(t)\right] \vec \Psi(R,t)
\label{schroedinger}
\end{equation}

\begin{figure}[hbt!]
\resizebox{0.45\textwidth}{!}{\includegraphics{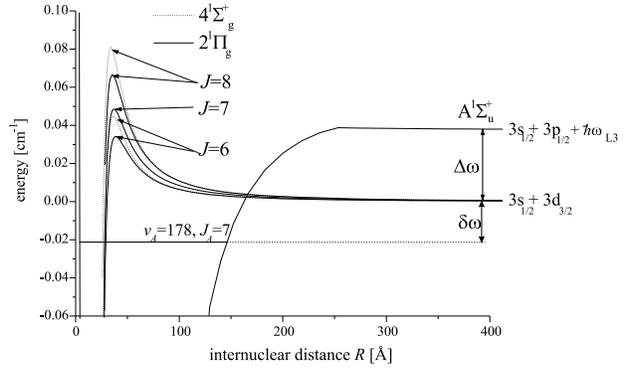}}
\caption{Asymptotic region of the coupled electronic potentials. The A state
potential is dressed by a photon energy of the coupling laser L3. The 
energy difference $\Delta\omega$ is due to the detuning of the coupling laser 
relative to the atomic transition. $\delta \omega$ is the relative position of 
one individual vibrational level $v_A$ to the 3s$_{1/2}$+3d$_{3/2}$ 
asymptote within the dressed state picture.}
\label{fig8}
\end{figure}

The radial wave functions for the nuclear motion in the electronic channels at 
the lower and upper asymptote are transformed via
\begin{eqnarray}
\label{xitrans}
\chi_{G}(R,t)=\tilde \chi_{G}(R,t) \exp (i \Delta\omega t),\nonumber\\
\chi_{\Sigma}(R,t)=\tilde \chi_{\Sigma}(R,t) \exp (-i \omega_0 t),\\ 
\chi_{\Pi}(R,t)=\tilde \chi_{\Pi}(R,t)\exp (-i \omega_0 t)\nonumber,
\end{eqnarray}
where $\omega_0 $ (see Figure~\ref{fig7}) is the transition frequency of the atomic 
\mbox{3$^2$p$_{1/2}$ $\rightarrow$ 3$^2$d$_{3/2}$}-transition, $\Delta\omega$ is 
the detuning of the coupling laser with respect to $\omega_0 $ 
($\Delta\omega= \omega_L - \omega_0 $). By the transformation (\ref{xitrans}) 
we shift (dress) the origin of energy for the electronic potentials,
\begin{eqnarray}
\tilde U_{G}(R)=U_{G}(R)+\hbar \Delta\omega\nonumber,\\
\tilde U_{\Sigma}(R)=U_{\Sigma}(R)-\hbar  \omega_0,\\
\tilde U_{\Pi}(R)=U_{\Pi}(R)-\hbar  \omega_0\nonumber.
\end{eqnarray}
Hence, all three electronic potentials have almost the same asymptotic energies. In Figure~\ref{fig8} the resulting molecular potentials dressed by one photon of a blue detuned (relative to the atomic transition) coupling laser are plotted.

Adapting the dressed potentials in the $\hat {\bf H}^{\rm mol}_i$, the Schr\"o\-dinger equation (\ref{schroedinger}) for the vibrational motion in the six coupled channels can be written as
\begin{equation}
\begin{split}
& i \hbar \frac{\partial }{\partial t}
\left(
\begin{array}{r}
\tilde \chi_G^J(R,t)\\
\tilde \chi_{\Sigma}^{J-1}(R,t)\\
\tilde \chi_{\Sigma}^{J+1}(R,t)\\
\tilde \chi_{\Pi}^{J-1}(R,t)\\
\tilde \chi_{\Pi}^{J}(R,t)\\
\tilde \chi_{\Pi}^{J+1}(R,t)\\
\end{array}
\right)=\\
& 
\left( 
\begin{array}{cccccc}
{\bf H}^{\rm mol}_{G_J}&\hbar {\bf \Omega}_{\Sigma}^{-1}&\hbar {\bf \Omega}_{\Sigma}^{+1}&\hbar {\bf \Omega}_{\Pi}^{-1}&\hbar {\bf \Omega}_{\Pi}^{0}&\hbar {\bf \Omega}_{\Pi}^{+1}\\
\hbar {\bf \Omega}_{\Sigma}^{-1}& {\bf H}^{\rm mol}_{\Sigma_{J-1}}&0&0&0&0\\
\hbar {\bf \Omega}_{\Sigma}^{+1}&0 & {\bf H}^{\rm mol}_{\Sigma_{J+1}}&0&0&0\\
\hbar {\bf \Omega}_{\Pi}^{-1}&0&0& {\bf H}^{\rm mol}_{\Pi_{J-1}}&0&0\\
\hbar {\bf \Omega}_{\Pi}^{0}&0&0&0& {\bf H}^{\rm mol}_{\Pi_{J}}&0\\
\hbar {\bf \Omega}_{\Pi}^{+1}&0&0&0&0& {\bf H}^{\rm mol}_{\Pi_{J-1}}\\
\end{array}
\right)
\left(
\begin{array}{r}
\tilde \chi_{G}^{J}(R,t)\\
\tilde \chi_{\Sigma}^{J-1}(R,t)\\
\tilde \chi_{\Sigma}^{J+1}(R,t)\\
\tilde \chi_{\Pi}^{J-1}(R,t)\\
\tilde \chi_{\Pi}^{J}(R,t)\\
\tilde \chi_{\Pi}^{J+1}(R,t)\\
\end{array}
\right),
\end{split}
\end{equation}
where $J$ is short for $J_A$ used in the other sections.

To solve this multi-channel problem the explicit time dependence of the Hamilton
operator can be eliminated if the rotating wave approximation \cite{Bws90} is 
applied. It is based on the suppression of the fast oscillating terms in the 
coupling matrix elements, which average out over  times given by the vibrational 
period of the molecule.

The coupling matrix elements between the $G$ (originally the A~state) and
electronically higher excited states $\Sigma$ and $\Pi$ are expressed by time 
independent Rabi frequencies ${\bf \Omega}_j^{\beta}(R)/ 2\pi$ 
\mbox{($j=\Sigma,\Pi$)} where \mbox{$\beta=0,\pm 1$} indicates the difference of 
the rotational quantum number between the coupled states.
They are related to the electronic transition dipole moment by 
\begin{eqnarray}
\hbar {\bf \Omega}_{j}^\beta(R)&=&- \;\;\frac{1}{2}\;\;\; E_0\;\;\; |\vec{D}_{G, j}(R)|
\left(
\begin{array}{ccc}
J\!\!+\!\beta & 1 & J\\
-M & 0 & M
\end{array}
\right)
\\
&=&- \frac{1}{2}\sqrt{\frac{2I_{\rm L3}}{c \epsilon_0}}\ |\vec{D}_{G, j}(R)|
\left(
\begin{array}{ccc}
J\!\!+\!\beta & 1 & J\\
-M & 0 & M
\end{array}
\right).\nonumber
\end{eqnarray}
Here we take into account the earlier discussed $M$-de\-pen\-dence by the
linear polarized light in the experiment. For each projection of $M$ on the 
laboratory frame the eigenvalue problem of the coupled channel system gives the 
stationary solution and can be solved for each $M$ separately. Levels with different 
$M$ do not couple.

The light induced energy shifts (the difference of corresponding eigenvalues
with and without light field) and line broadenings are computed using the Mapped 
Fourier Grid Hamiltonian (MFGH) representation, described in detail in 
\cite{Koko99}. The MFGH method has been designed in the context of 
photoassociation of cold atoms and cold molecule formation, where interactions 
at large internuclear distances play the key role. In subsequent papers 
\cite{Koko00,pell02}, a complex potential has been introduced to treat the 
interaction of bound levels embedded into dissociation continua. Here we recall 
only the main steps of the MFGH method:

\begin{itemize}
\item In the standard Fourier Grid Hamiltonian (FGH) method, the total
Hamiltonian for the coupled channels is expressed in a basis on $N$ plane waves, 
$\exp\bigg({\frac{i2\pi kR}{L}}\bigg)$, 
$k=-(\frac{N}{2}-1),\dots,0,\dots,\frac{N}{2}$, where $L$ is the range of 
internuclear distances under investigation. The diagonalization yields 
eigenfunctions represented as an expansion over the same basis, the coefficients 
being the value of the wave function at each point of the grid of length $L$ 
with $N$ equi-distant points.

\item As the local de Broglie wavelength of wave functions varies by orders of
magnitude between the small and the large internuclear distance range, an 
adaptative coordinate can be defined in the FGH framework, which maps the 
variation of the local kinetic energy. The number of required grid points is 
reduced substantially, allowing accurate calculations of energy levels with 
large elongations, i.e., close to the dissociation limit.

\item The width of the energy levels interacting with a dissociation continuum
is 
obtained after the diagonalization of the total Hamiltonian, including a pure 
imaginary potential (or an optical potential) $V_{\rm opt}$ placed at large 
distances, which ensures absorbing boundary conditions for the outgoing waves. 
In a stationary approach, the dissociating wave functions behave asymptotically 
as Siegert states \cite{pell02}.
We choose the form proposed in \cite{Vib92}:

\begin{equation}
\label{vopt}
V_{\rm opt}=A_{\rm opt} N_{\rm opt}exp\bigg(\frac{-2L_{\rm opt}}{R-R_{\rm opt}}\bigg),
\end{equation}
where $N_{\rm opt}$ is a normalization factor, $R_{\rm opt}$, $L_{\rm opt}$ and $A_{\rm opt}$ 
are
respectively the position, the length, and the amplitude of the optical 
potential. The values of these parameters are chosen according to the 
recommendation of \cite{Vib92}.
The imaginary part of the eigenvalues resulting from the diagonalization of the
Hamiltonian matrix provides the predissociation line width
\end{itemize}

\noindent In our calculations, we used a grid with 
$L\:$=1100~a.u.(1~a.u. =$0.5291772083 \times 10^{-10}$~m), with typically 
700 grid points, which induces a diagonalization of 4200$\times$4200 squar\-ed 
matrix, using standard \mbox{LAPACK} routines. As we are looking for very small energy 
shifts, a high accuracy is needed for the eigenvalues, which is checked by 
studying their convergence depending on the contraction (equal to 0.2 here) of 
the grid step. A contraction of 1 corresponds to a local grid density of four 
points per local de Broglie wavelength. The optical potential is set up to 
absorb outgoing waves with kinetic energies corresponding to the
detuning $\Delta\omega$ of the coupling laser L3, following the rules described 
in \cite{Vib92}. The parameter values are $A_{\rm opt}=5.6 \times
10^{-7}$~a.u., $L_{\rm opt}=210$~a.u., $R_{\rm opt}=L-L_{\rm opt}$, and
$N_{\rm opt}=13.22$. Changes of $A_{\rm opt}$, $L$
and $L_{\rm opt}$ in reasonable ranges do not affect the simulation results for
the line positions and the line widths, so that their choice is appropriate to
represent the loss of molecules by predissociation.

For the simulations, we applied the following potentials: For the
$A^1\Sigma^+_u$~state a RKR potential (Rydberg-Klein-Rees) generated from 
experimental values is used. It is based on energies from \cite{Tie96}. The 
potentials for $4^1\Sigma^+_g$ and $2^1\Pi_g$~state for internuclear distances 
up to 35 a.u. are taken from {\it ab~initio} calculations in \cite{Mag93} 
(method b) therein). The long range part is calculated from the dispersion 
coefficients given in \cite{Mar95} and is connected continuously differentiable 
to the inner part.

The dipole moments used for the light induced coupling originate 
from calculations in \cite{Mag93}. They converge for internuclear distances above 100 a.u.
to the atomic values. For the inner part below 14 a.u., we generated 
additional values of the dipole moment by smooth continuation of the {\it 
ab~initio} curves. This inner part of the dipole moments has only minor 
influence  on the results of the numerical simulation as the contribution to the 
overlap integral of the A~state and upper states wave functions at small 
internuclear distances is almost negligible.

\begin{figure}
\resizebox{0.45\textwidth}{!}{\includegraphics{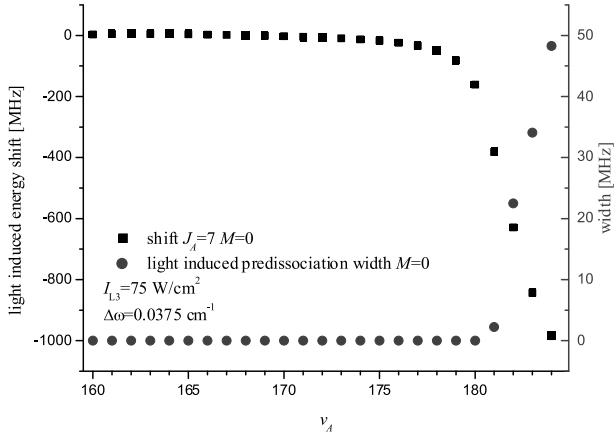}}
\caption{Simulation of light shifts and line broadenings of the last 25 vibrational levels of the $A^1\Sigma^+_u$~state calculated for $J_A\:$=7 and an intensity of the coupling laser of 75~W/cm$^2$. The coupling laser is assumed to be 0.0375~cm$^{-1}$ blue detuned to the atomic \mbox{3$^2$p$_{1/2}\rightarrow$~3$^2$d$_{3/2}$} transition. All levels of the A state are shifted to lower energies.}
\label{fig9}
\end{figure}

For the simulation of line shifts of asymptotic A~state levels, we compare the
theoretical energy positions of the single channel calculation of the A~state 
with the corresponding values of the six channel system. A typical result of a 
simulation is presented in Figure~\ref{fig9}. We display the energy 
shifts (squares) of vibrational levels $v_A=165,\dots,184$ with $J_A$=7 for a peak 
intensity of the coupling laser $I_{\rm L3}\:$= 75 W/cm$^2$ and a blue detuning of 
0.0375~cm$^{-1}$ with respect to the atomic  
\mbox{3$^2$P$_{1/2}\rightarrow$~3$^2$D$_{3/2}$} transition. Moreover, the figure 
contains light induced line broadenings (circles) obtained from the simulations with 
imaginary absorbing potential. With the blue detuning of 0.0375 cm$^{-1}$ all 
levels bound by less than this energy starting with $v_A=181$ show 
considerable predissociation broadening of up to 49~MHz.

\begin{figure}
\resizebox{0.45\textwidth}{!}{\includegraphics{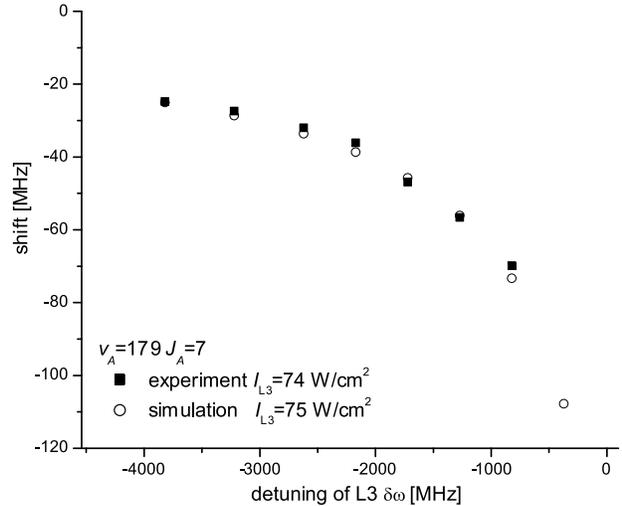}}
\caption{Dependence of the light shift of $v_A\:$=179,~$J_A\:$=7 on the 
detuning of the coupling laser relative to a hypothetical transition
\mbox{($v_A\:$=179,~$J_A$=7)$\rightarrow$3s$_{1/2}$+3d$_{3/2}$} asymptote (see 
also Figure~\ref{fig7} and Figure~\ref{fig8}). The squares are 
experimental points taken for a peak intensity of $I_{\rm L3}\:$=74~W/cm$^2$. 
The circles are from a simulation with $I_{\rm L3}\:$=75~W/cm$^2$.}
\label{fig10}
\end{figure}

Similar calculations have been performed for all studied laser intensitites
$I_{\rm L3}$ and detunings $\Delta\omega$ of the coupling laser. In 
\mbox{Figure \ref{fig10}} the circles show a simulation of the dependence of level shifts for $v_A\:$=179, $J_A\:$=7 on the detuning 
$\delta\omega$ of the coupling laser  with respect to a hypothetical transition 
\mbox{($v_A\:$=179,~$J_A\:$=7)$\rightarrow$3s$_{1/2}$+3d$_{3/2}$} asymptote. 
Experimental points are plotted by squares.

\begin{figure}
\resizebox{0.45\textwidth}{!}{\includegraphics{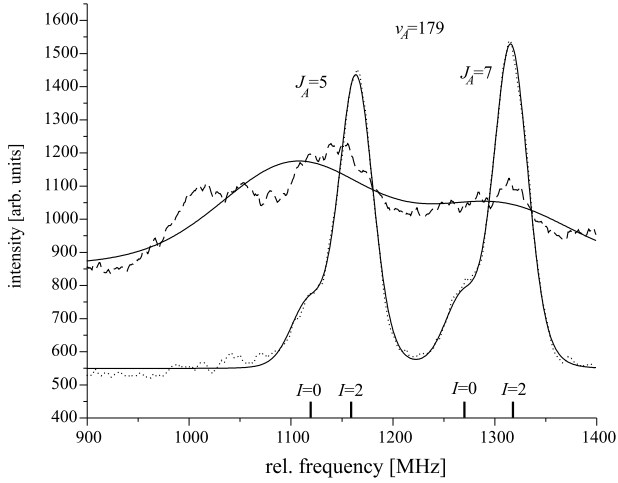}}
\caption{Example of an experimental scan of $v_A\:$=179,~$J_A\:$=5,7 
when the coupling laser is on resonance ($\delta\omega=0$) with respect to an hypothetical 
transition  \mbox{$v_A\:$=179,~$J_A\:$=7$\rightarrow$3s$_{1/2}$+3d$_{3/2}$} 
asymptote. The intensity of L3 was $I_{\rm L3}$=74 W/cm$^2$. The dotted trace 
is taken without the coupling laser, while the dashed one is the manipulated 
spectrum. The solid traces represent fits on the experimental data with the 
model presented in Section~\ref{fitdata}.}
\label{fig11}
\end{figure}

The agreement between experimental  and simulated line shifts 
is convincing, but one has to keep in mind that due to the uncertainty in the
determination of the coupling laser beam waist, we have a 50\% uncertainty 
in the laser intensities applied for the simulation.
For detuning less than 1000~MHz our fitting procedure fails to determine line
shifts from the experimental spectra. The experimental signal of the shifted 
trace decreases and broadens remarkably towards the resonance frequency. A 
typical spectrum showing this effect for $v_A=179$, $J_A=5, 7$ is 
presented in Figure~\ref{fig11}. The theoretical model applied for the fits of 
line profiles is no longer able to simulate the observed line profiles properly. 

The deviations in Figure~\ref{fig11} are not noise; the slow variation is reproducible.
 For a proper simulation, the fit should include the fine and hyperfine structure 
on the optical coupling.

\begin{figure}[htb!]
\resizebox{0.45\textwidth}{!}{\includegraphics{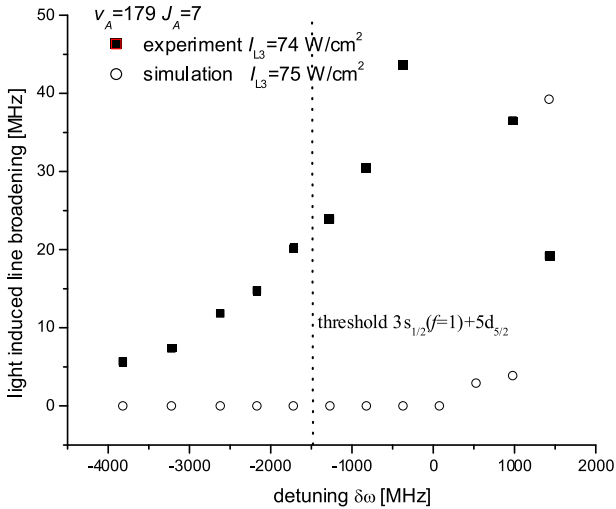}}
\caption{Dependence of the line broadening of $v_A=179$, $J_A=7$ on the
detuning  $\delta \omega$  of the coupling laser (see Figure~\ref{fig8}). 
The squares represent the difference of the two FWHM fitting parameters. The 
circles are from a simulation for the experimental peak intensity of $I_{\rm 
L3}\:$=74~W/cm$^2$, neglecting the fine and hyperfine structure of the asymptote \mbox{3s+3d}. The threshold is only given for the discussion in the text.}
\label{fig12}
\end{figure}

A comparison of the line broadening observed in the experiment
with theoretically predicted line broadening due to laser induced predissociation 
is shown in Figure~\ref{fig12}. The experimental broadenings are 
calculated from the fitted FWHM in presence of the coupling laser minus the FWHM 
without light induced broadening. In the experiment we observe a significant line 
broadening, which has a maximum if the coupling laser is on resonance with a 
hypothetical transition frequency from the level under investigation to the 
\mbox{3s$_{1/2}$+3d$_{3/2}$} asymptote ($\delta\omega=0$). The broadening 
effect for $\delta\omega < 0$ cannot be described by the laser induced 
predissociation calculated with our six channel model because the dissociation 
channel is not yet open.

\begin{figure}
\resizebox{0.45\textwidth}{!}{\includegraphics{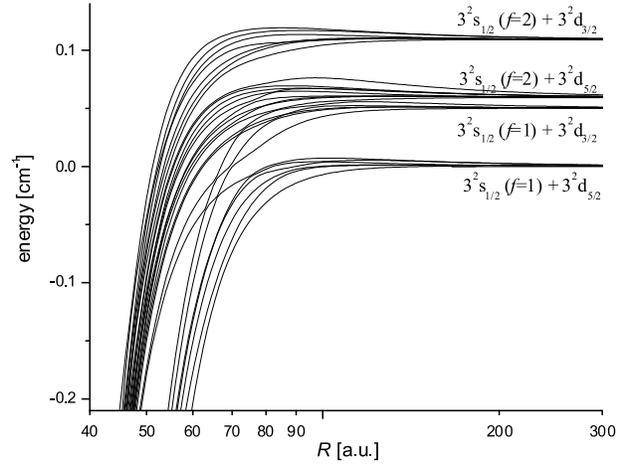}}
\caption{Hyperfine potentials at the asymptote $3s+3d$ for a total angular
momentum $F=0$.}
\label{fig13}
\end{figure}

There are two different effects, which can lead to the experimentally observed
line broadening for detuning $\delta\omega < 0$. The molecular structure of the 
states at the 3s+3d asymptote is not as simple as we have assumed for the 
simulation. Due to the fine structure energy of the 3d atom and hyperfine energy 
of the 3s atom, which have almost the same magnitude ($E_{\rm FS}$(3d)=-0.0498 
cm$^{-1}$ \cite{Bur87}, $E_{\rm HFS}$(3s)=-0.0590 cm$^{-1}$ \cite{Ar77}) the 
3s+3d asymptote splits into four different hyperfine asymptotes. The calculated 
adiabatic asymptotic potentials are shown in Figure \ref{fig13}. From the A 
state, which correlates to the \mbox{3s$_{1/2}(f=1)$+3p$_{1/2}$} asymptote, a 
dipole transition is only possible to the \newline 
\mbox{3s$_{1/2}(f=1)$+3d$_{3/2}$} asymptote. Thus, the light induced shifts of 
an asymptotic A~state level shows a frequency dependence which can be described by an 
effective coupling to only this asymptote. In the region 
above the \mbox{3s$_{1/2}(f=1)$+3d$_{5/2}$} asymptpote, which is the lowest 
asymptote due to the inverted fine structure, a predissociation can appear. This will lead 
to an increased line width of the A state levels under investigation starting at 
a detuning $\delta\omega > -|E_{\rm FS}|$. This threshold is included in Figure~\ref{fig12}.

\begin{figure}
\resizebox{0.45\textwidth}{!}{\includegraphics{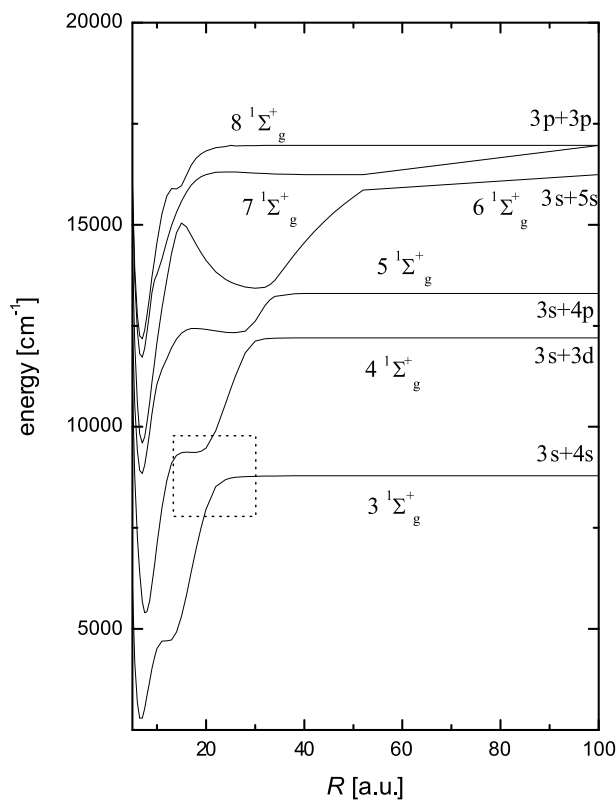}}
\caption{Molecular potentials with $^1\Sigma^+_g$-symmetry. The shape of the  
4$^1\Sigma^+_g$-potential is due to an avoided crossing with the 
3$^1\Sigma^+_g$-state that correlates to the $3s+4s$ asympote \cite{Mag93}.}
\label{fig14}
\end{figure}

A second effect leading to a predissociation and thus a decreased lifetime of
the 4$^1\Sigma^+_g$ state is the vibrational coupling of this state to lower
lying 
molecular states with the same symmetry. In Figure~\ref{fig14} the region of an 
avoided crossing with the 3$^1\Sigma^+_g$-state is marked by a dashed box. The 
3$^1\Sigma^+_g$ state correlates to the 3s+4s asmyptote so that in the region 
above this asymptote a predissociation to this asymptote is possible. 
Therefore, the optical coupling of the A~state to the 4$^1\Sigma^+_g$-state can also induce a predissociation of A~state levels to the 3s+4s asymptote.

Due to limitations of the memory available on our computers the two additional
predissociation channels can not be implemented in our coupled channel code. 
Hence, it is difficult to estimate the contribution of the two effects to the 
experimentally observed line width.

\section{Conclusion}
A high resolution molecular beam experiment has been used to investigate the effect of a light induced 
coupling on the last vibrational levels of the sodium A~state. The coupling is induced by a laser field 
which is near resonant to the atomic sodium 3$^2$P$_{1/2}\rightarrow 3^2$D$_{3/2}$ transition and thus 
couples the $A^1\Sigma^+_u$ state to the $4^1\Sigma^+_g$ and $2^1\Pi_g$ state, which are both correlated 
to the \mbox{3s$_{1/2}$+3d$_{3/2}$} asymptote. From the experimental data we see that with the usual 
laser intensities available from lab size laser sources ($\approx~50\;$W/cm$^2$) light induced level 
shifts in a molecular system in the order of several 10 MHz can be induced. Furthermore, we observe 
an increased line width of the A~state levels under investigation due to light induced predissociation. 
The number of bound states of the A state could be changed by the influence of the coupling laser by one or more units. Thus, in the picture of colliding atoms, the scattering phase was altered by more than $\pi$.

We developed a theoretical model of the light induced coupling, which describes the observed level 
shifts in good agreement with the experimental data. It is based on a six channel calculation using 
the Mapped Fourier Grid Method for the solution of the coupled channel eigenvalue problem. We use 
{\it ab initio} data for the electronic potentials at the \mbox{3s$_{1/2}$+3d$_{3/2}$} asymptote and 
for the dipole moments for the description of the light induced coupling. For the simulation of the 
induced line shifts of deeply bound states it is not necessary to take into account the fine and 
hyperfine structure of the 3s+3d asymptote. 

The interpretation of the light induced line broadening effects needs a more detailed analysis of 
the molecular structure at the 3s+3d asymptote. The predissociation of the $4^1\Sigma^+_g$ and 
$2^1\Pi_g$ state due to fine structure, hyperfine structure, and vibrational motion must be taken 
into account. 

This research is in progress with a similar experiment that couples the sodium ground state asymptote with the 3s+3p asymptote \cite{sam03}. Here we want to investigate systematically a laser induced coupling between the 
\mbox{X$^1\Sigma^+_g$} ground state of sodium to levels in the A state. This experimental situation 
is of very high interest, as in most cooling and trapping experiments lasers (either for a MOT or 
dipole trap) are present, which are slightly detuned from the atomic resonance. These experiments show very complex dynamical Stark effect which asks for further extensions of our model system.

\section{Acknowledgments}
We thank the Deutsche Forschungsgemeinschaft supporting this work within the \mbox{SFB 407}. We also 
thank the \mbox{PROCOPE} program, which provided the money for the travelling expenses between France and 
Germany.

 \bibliographystyle{}
 \bibliography{}
%
 
 
 

\end{document}